\def\up{\uparrow}
\def\dn{\downarrow}
\def\ts{\thinspace}
\def\amp{{\cal A}}
      [1999/12/01 v1.4c Il Nuovo Cimento]
\documentclass{cimento}

\usepackage{graphicx}  
\title{Spin correlations: Tevatron {\it vs.}\ LHC}
\author{Gregory~Mahlon\from{ins:x}}
\instlist{\inst{ins:x} Penn State Mont Alto, 1 Campus Drive,
Mont Alto, PA 17237 U.S.A.}
\PACSes{\PACSit{14.65.Ha}{Top quarks}}
\begin{document}

\maketitle

\begin{abstract}
We compare theoretical expectations for the observation of
spin correlations in top quark pair production and decay at
the Fermilab Tevatron and the CERN Large Hadron Collider (LHC).
In particular, we note that the differing top quark pair production
mechanisms in the two environments test different 
aspects of the Standard Model and
require different strategies to observe the correlations.
At the Tevatron, production is dominated by $q\bar{q}\rightarrow t\bar{t}$
and the strategy is to construct a double-decay angle
distribution where one decay angle is measured in the $t$
rest frame and the other in the $\bar{t}$ rest frame.
The dominant process at the LHC is $gg\rightarrow t\bar{t}$,
with a rich spin structure that allows for a second option in
observing spin correlations.  Here the strategy is to select
events where the $t\bar{t}$ pair is produced at relatively low
velocity in the zero momentum frame (ZMF).  For these events, there are
strong azimuthal correlations between $t$ and $\bar{t}$ decay products
present.  This measurement enjoys the advantage that it
can be carried out in the laboratory frame.
\end{abstract}

\section{Introduction}
One of the major components of the physics program at the LHC 
will be a detailed study of the properties of the top quark.
In this talk, we report on recent theoretical work on how to
observe spin correlations in the production and decay of top
quark pairs via gluon fusion\cite{ref:MPlhctop}.
The top quark is unique among the quarks in the Standard Model
in that its decay lifetime is so short that it is predicted to
decay before spin decorrelation takes place\cite{ref:Bigi}.
Thus, top quarks produced in a particular spin state will pass
information about that state on to its decay products,
and an investigation of the angular correlations among the decay
products in select $t\bar{t}$ events can
shed light on the details of the production and decay mechanisms.
The mere observation of $t\bar{t}$ spin correlations would provide
an upper limit on the top quark lifetime; conversely, the observation
of a lack of significant $t\bar{t}$ spin correlations would indicate
the participation of the top quark in some sort of new strong 
interaction that causes spin decorrelation to occur more quickly than
predicted by the Standard Model considerations of Ref.~\cite{ref:Bigi}.
Thus, in the near-term, establishing the presence or absence of
these correlations is a worthwhile physics goal.  In the longer term,
a detailed study of as many aspects of the correlations as possible
will provide a series of additional wide-ranging tests of the
Standard Model.  

The outline of 
this talk is as follows.  We will begin
with a detailed examination of the spin structure of the main
top pair production mechanisms at the Tevatron  
($q\bar{q}\rightarrow t\bar{t}$) and LHC ($gg\rightarrow t \bar{t}$),
with a particular emphasis on elucidating the spin state(s) of the
final $t\bar{t}$ pairs.  In Sec.~\ref{sec:decays} we will examine the 
angular distributions associated with the decay of polarized top
quarks.  Combining the production properties with the decays will
lead us to the strategies for observing the spin correlations.
These strategies will be different for the Tevatron (Sec.~\ref{sec:FNAL})
and LHC (Sec.~\ref{sec:LHC}).
In Sec.~\ref{sec:future} we will briefly consider 
some of the physics that could
be accomplished with ${\cal O}(10^6)$ $t\bar{t}$ pairs at our
disposal.  Finally, Sec.~\ref{sec:conclusions} contains a summary 
and our conclusions.  Additional details 
may be found in Ref.~\cite{ref:MPlhctop}.

\section{Detailed spin structure in top quark pair production}

The main two top quark pair production mechanisms at hadron colliders
are $q\bar{q}\rightarrow t\bar{t}$, which dominates
at the Fermilab Tevatron, and $gg\rightarrow t\bar{t}$, which
dominates at the LHC.  In this section, we examine the spin 
structure of each of these processes in some detail.

The process $q\bar{q}\rightarrow t\bar{t}$ proceeds through an
$s$-channel gluon; in order to couple to this gluon, the incoming
$q\bar{q}$ pair must possess opposite helicities.  The initial state
is thus either $q_R \bar{q}_L$ or $q_L \bar{q}_R$.  We will 
focus our discussion on $q_R \bar{q}_L$:  results for $q_L \bar{q}_R$
may be generated by flipping all of the spins in the initial and
final states.

\begin{figure}[hbt]
\hfil\includegraphics[height=1.6truein,keepaspectratio=true]{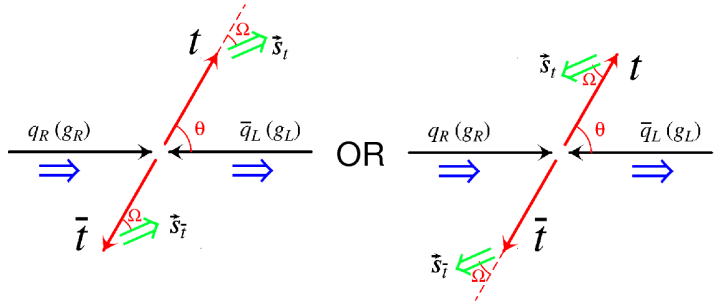}
\caption[]{The spin configurations for the process 
$q_R \bar{q}_L \rightarrow t\bar{t}$ are best described by the
off-diagonal basis.  The angle between the top quark momentum and
the spin vector in the zero momentum frame 
is given by $\tan\Omega = (1-\beta^2)\tan\theta$
where $\beta$ is the speed of the top quark in the ZMF.
The ratio of $t_{\up} \bar{t}_{\dn}$ to $t_{\dn} \bar{t}_{\up}$
production is 
$(1+\sqrt{1-\beta^2\sin^2\theta})^2:(1-\sqrt{1-\beta^2\sin^2\theta})^2$.
The results for the $q_L \bar{q}_R$ initial state may be obtained
by reversing all of the spins in the diagram.  The same spin 
structure also applies to $gg\rightarrow t\bar{t}$ with
opposite helicity gluons.}
\label{fig:qqspins}
\end{figure}

The top quark pairs produced from the $q_R \bar{q}_L$ are well-described
by the off-diagonal spin basis\cite{ref:OD} 
(see Fig.~\ref{fig:qqspins}):  
at leading
order the top quarks have opposite spins ($t_\up\bar{t}_\dn$ or
$t_\dn\bar{t}_\up$) 100\% of the time using this 
spin quantization axis.  
At threshold, the off-diagonal basis coincides with
the beamline basis\cite{ref:MP1}, the appropriate
choice of quantization axis in the $\beta\rightarrow0$ 
limit\cite{ref:thinkshop}.  At the other extreme ($\beta\rightarrow1$),
the off-diagonal basis coincides with the helicity basis.
Between these two extremes, the
off-diagonal basis smoothly interpolates
from the beamline basis at small $\beta$ to the 
helicity basis in the ultra-relativistic
limit while maintaining a 
spin state consisting of 100\% opposite spin
$t\bar{t}$ pairs.
\noindent
\noindent


The gluon fusion top pair production mechanism possesses a
very rich spin structure:  the spin state of the $t\bar{t}$ is
different for different gluon helicities. 
When the initial gluons have opposite
helicity, the $t\bar{t}$
spin state is the same as for $q\bar{q}$ production
(Fig.~\ref{fig:qqspins}).  
Thus, the off-diagonal basis provides the description with
the maximum $t\bar{t}$ spin correlation in this case.
Opposite helicity gluon pairs form the
dominant contribution to $gg\rightarrow t\bar{t}$
when $\beta\gamma\sin\theta > 1$,
or, equivalently, when $\beta^2 > 1/(2-\cos^2\theta)$.

On the other hand, when $\beta\gamma\sin\theta <1$, like-helicity
gluon pairs form the dominant contribution.
In this case, it turns out that the helicity basis provides the
best description of the $t\bar{t}$ spin correlations, 
no matter
what the value of $\beta$.  
Like helicity gluons produce like-helicity $t\bar{t}$ pairs.
The ratio of $t_R\bar{t}_R$ to $t_L\bar{t}_L$ production from
a $g_R g_R$ initial state is given by $(1+\beta)^2:(1-\beta)^2$.


\section{Decay of polarized top quarks}\label{sec:decays}

We now consider the angular distributions associated
with the decay of spin up top quarks.  We define the decay angles
in the top quark rest frame; 
we will denote the angle between the $i$th particle and the top
quark spin direction by $\theta_i$.  Then, the decay angular
distribution takes the simple form\cite{ref:LOalpha}
\begin{equation}
{{1}\over{\Gamma}}
{{\drm\Gamma}\over{\drm(\cos\theta_i)}}
= {{1}\over{2}}( 1 + \alpha_i \cos\theta_i).
\end{equation}
The analyzing power $\alpha_i$ depends on which decay product we
consider; the values of $\alpha_i$ at next-to-leading 
order\cite{ref:NLOalpha}
are collected in Table~\ref{tab:alphas}.

\begin{table}
  \caption{Analyzing powers $\alpha$ for both semi-leptonic 
and hadronic top quark decays at next-to-leading order.
The coefficients for $u,d,s,c$ and $b$ are for partons: the values
for jets differ slightly at NLO (see Ref.~\cite{ref:NLOalpha}).}
  \label{tab:alphas}
  \begin{narrowtabular}{2cm}{ccrc}
    \hline
   &   Decay product   & $\alpha\quad$ &  \\
    \hline
   &   $\ell^{+}$         & $0.998$  &  \\
   &   $\bar{d}, \bar{s}$ & $0.966$  &  \\
   &   $\nu$              & $-0.314$ &  \\
   &   $u, c$             & $-0.317$ &  \\
   &   $b$                & $-0.393$ &  \\
    \hline
  \end{narrowtabular}
\end{table}

The normalized distribution
for the complete $2\rightarrow6$ production and decay process
is
\begin{equation}
{ {1}\over{\sigma} }
{ 
{\drm^2\sigma}
\over
{ \drm(\cos\theta_i)
  \drm(\cos\bar\theta_{\bar\imath}})
}
= {{1}\over{4}}
\Bigl[
1 + {{ N_\parallel - N_\times }\over{N_\parallel+N_\times}}
\alpha_i \bar\alpha_{\bar\imath} 
\cos\theta_i \cos\bar\theta_{\bar\imath}
\Bigr].
\label{eq:doublediff}
\end{equation}
Eq.~(\ref{eq:doublediff}) clearly displays the dependence on 
production ($N_\parallel$ is the number of events with
like-spin $t\bar{t}$ pairs; $N_{\times}$ is the number of
events with opposite-spin $t\bar{t}$ pairs) and decay
($\theta_i,\alpha_i$ refer to the $t$ side of the event;
$\bar\theta_{\bar\imath},\bar\alpha_{\bar\imath}$ refer to
the $\bar{t}$ side of the event; the angles are measured
in the respective rest frames of the $t$ and $\bar{t}$\ts).
The next-to-leading-order corrections to this distribution
have been presented in Refs.~\cite{ref:NLOdouble1,ref:NLOdouble2}; 
these corrections will play an important 
role in future precision measurements
of the correlations.

\section{Tevatron strategy for observing spin correlations}\label{sec:FNAL}

At the Fermilab Tevatron, the strategy for observing the spin
correlations in $t\bar{t}$ production and decay 
consists of reconstructing 
the double decay angular distribution 
Eq.~(\ref{eq:doublediff}) from the data.
An examination of Eq.~(\ref{eq:doublediff}) indicates how to maximize
the size of the correlations.
First, one should
choose a spin quantization axis that maximizes the asymmetry
$(N_{\parallel}-N_{\times})/(N_{\parallel}+N_{\times})$.  Since the
dominant process at the Tevatron is $q\bar{q}\rightarrow t \bar{t}$,
this means using the off-diagonal basis, although in the long
term it will be worthwhile to do the measurement using the helicity
and beamline bases as well.  Measurements with multiple spin bases
probe different aspects of the $q\bar{q}\rightarrow t\bar{t}$ matrix
element and provide a cross-check on the analysis.  Furthermore, the
sources of systematic uncertainty are likely to vary greatly
for different spin bases. 

Second, one should choose $t$ and $\bar{t}$ decay products
with large analyzing powers (\idest\ the charged lepton or $d$-type
quark).   This consideration suggests looking in the dilepton 
or lepton+jets modes.  The dilepton mode has the advantage of the
maximum possible analyzing power, but suffers from a small branching
ratio and the complications associated with the two (unobserved)
neutrinos in the final state.  The lepton+jets mode couples higher
statistics with a much better ability to reconstruct the $t$ and
$\bar{t}$ rest frames, but has a lower analyzing power since the
jet associated with the
$d$-type quark must be selected probabilistically.
Eventually, given sufficient statistics, it will be
useful to measure the correlations for as many different combinations
of decay products as possible.

Although this is a very difficult analysis, both CDF
and D0 have reported initial attempts to observe
these correlations  
despite the limited data set currently available.
At present, with only about half of the approximately 7~fb$^{-1}$
delivered by the Tevatron analyzed so far, no unambiguous ($>3\sigma$)
sign of the correlations has been seen; however, both CDF and D0
plan on updates as well as a combined result 
in the near future\cite{ref:headproc}.

\section{LHC strategy for observing spin correlations}\label{sec:LHC}

The dominant top quark pair production mechanism at the LHC is
gluon fusion.  The detailed study of this process using an arbitrary
spin axis performed in Ref.~\cite{ref:MPlhctop} leads us to the
following two conclusions:
When $\beta\gamma\sin\theta < 1$, the best
spin basis is the one which maximizes the like-spin fraction
(like-helicity gluons dominate in this region).
On the other hand, when $\beta\gamma\sin\theta>1$, the best
spin basis is the one which maximizes the opposite-spin fraction
(opposite-helicity gluons dominate in this region).  
%

Because the LHC will be a copious source of $t\bar{t}$ pairs
(about $10^6$ $t\bar{t}$ pairs per fb$^{-1}$ at full beam
energy; even at reduced energy there will be $10^4$ to $10^5$
pairs per fb$^{-1}$), there will be room to implement significant
cuts on the data before statistical uncertainties become 
comparable to the systematic uncertainties.  Thus, we will focus
on the $\beta\gamma\sin\theta < 1$ region.  In this region,
like-helicity gluons to like-helicity $t\bar{t}$ pairs dominate.
Since $\beta$ is restricted to fairly moderate values in most of this
region, the spin correlations won't be masked by large boosts, and
one could, in principle, employ the same analysis as has been used
at the Tevatron.  Another option exists, however.

In order to motivate this alternative, we will examine the ratio
\begin{equation}
{\cal S}  \equiv
{
{(\vert\amp\vert^2_{RR} + \vert\amp\vert^2_{LL})_{\rm corr}}
\over
{(\vert\amp\vert^2_{RR} + \vert\amp\vert^2_{LL})_{\rm uncorr}}
}
\end{equation}
which compares the sum of the squares of the like-helicity amplitudes
for the fully-correlated $gg\rightarrow t\bar{t}$ matrix 
elements to the same sum for matrix elements calculated with
a toy model employing top quark decays which are spherical
in their rest frames (but with all other decays -- \idest\
the $W$'s -- fully correlated).  In terms of the cosines of the
angles in the ZMF between various pairs of particles and the ZMF
speed of the top quarks, 
\begin{equation}
{\cal S}  =
\Biggl[ {{1-\beta^2}\over{1+\beta^2}} \Biggr]
\Biggl[
{
{ (1+\beta^2)
  +(1-\beta^2)c_{\bar{e}\mu}
  -2\beta^2 c_{t\bar{e}} c_{\bar{t}\mu} }
\over
{ (1-\beta c_{t\bar{e}})
  (1-\beta c_{\bar{t}\mu}}) }
\Biggr],
\end{equation}
which reduces to $1+c_{\bar{e}\mu}$ for $\beta\rightarrow 0$.
So, at smallish values of $\beta$, the difference between the 
correlated and uncorrelated versions of the matrix elements is
sensitive to the angle between the two charged leptons, suggesting
that we examine $\Delta\phi$, $\Delta\eta$, and $\Delta R$ for
these two particles.   Of these three distributions, $\Delta\phi$
turns out to be the most sensitive to the presence or absence of
$t\bar{t}$ spin correlations.  This distribution also enjoys the
advantage of being invariant under longitudinal boosts; once
the event sample has been selected, this
measurement can be made in the laboratory frame without the
complications associated with the reconstruction of and boost
to some special frame of reference\footnote{Unfortunately, the
$\Delta\phi$ distribution in $q\bar{q}\rightarrow t\bar{t}$ is 
insensitive to the presence or absence of spin correlations.  Thus, 
this observable is not particularly interesting at the Tevatron.}.

\begin{figure}[hbt]
\hfil\includegraphics[height=1.9truein,keepaspectratio=true]{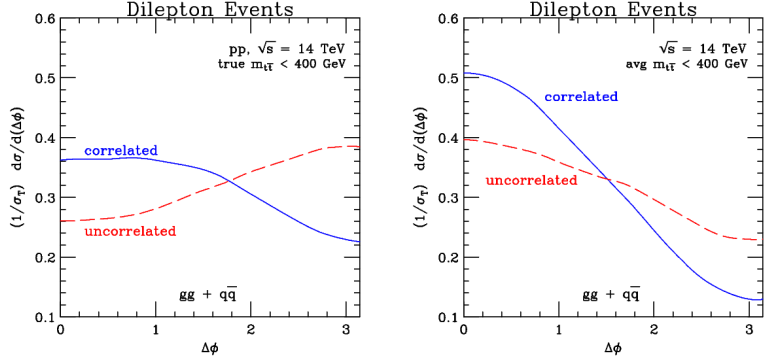}
\caption[]{The differential distribution in $\Delta\phi$,
$(1/\sigma_{{}_T}) \drm\sigma/\drm(\Delta\phi)$.
The solid curve is for the fully correlated case whereas the
dashed curve assumes that the top quarks decay spherically in their
respective rest frames.  In the plot on the left, a cut restricting
the (true) invariant mass of the $t\bar{t}$ pairs to a maximum of
400~GeV has been applied to the distributions; in the plot on the
right, the cut restricts the (na{\"\i}ve) unweighted average
of all of the reconstructed values of $m_{t\bar{t}}$ for each event
to a maximum of 400~GeV.}
\label{fig:dphi}
\end{figure}

In Fig.~\ref{fig:dphi}, we present a comparison of the correlated
and uncorrelated distributions in $\Delta\phi$ for the two
leptons in a sample of dilepton events.  In the plot on the left,
the events were required to have a $t\bar{t}$ invariant mass
of less than 400 GeV to ensure that they sample 
the like-helicity-gluon-dominated $\beta\gamma\sin\theta<1$
region of phase space.  Clearly this distribution has significant
sensitivity to the presence or absence of spin correlations.
Unfortunately, the pair of neutrinos in these events makes a
full reconstruction of the event 
impossible\footnote{Although there are a total of 8 constraint equations
for the 8 unknown components of the $\nu$ and $\bar\nu$ 4-momenta, two
of these constraints are quadratic, leading to up to 4 distinct 
solutions for each of the two possible pairings of 
$W$-bosons (leptons) with $b$ jets.  Thus, there are up to 8 distinct
values of $m_{t\bar{t}}$ generated by the reconstruction program.},
and we must look for
an alternate means of selecting the $\beta\gamma\sin\theta<1$ region.
In the plot on the right, the event selection is based on the
value of the (na{\"\i}ve) unweighted average of the 
various $m_{t\bar{t}}$ values produced by the neutrino reconstruction 
routine.
Cutting on this quantity has the unfortunate side-effect of producing
a systematic depletion of events near $\Delta\phi=\pi$; however,
this depletion affects both models in a similar fashion and
significant discriminating power remains.  What we have here is a
proof-of-concept:  to do the actual measurement it will be necessary
to understand the detector systematics and NLO corrections to
the $\Delta\phi$ distribution very well.  Left open are the questions of
whether a better substitute for the true value of $m_{t\bar{t}}$
exists as well as the optimal maximum value of $m_{t\bar{t}}$ to use
in selecting the data.

\noindent

We now turn to the lepton+jets mode.   This channel has the advantage
of larger statistics than dilepton mode, and only a single, 
over-constrained, neutrino to be reconstructed.  Thus, it is
possible to do a good job of locating the ZMF and calculating
$m_{t\bar{t}}$.  This mode has one disadvantage, however:  it is not
possible to know with 100\% certainty which of the two $W$-decay jets
corresponds to the $d$-type quark.  Thus, it is necessary to use
our best informed guess to select this jet on a probabilistic basis.
The authors of Ref.~\cite{ref:MP1} suggest using the jet which has the
smallest spatial separation from the $b$ jet in the $W$ rest frame;
this is equivalent to selecting the jet with the lowest energy in the
$t$ rest frame\cite{ref:denergy}.  Although this is the correct 
choice only about 60\% of the time, it is good enough to provide
an analyzing power of 0.4734 to 0.4736 at NLO\cite{ref:NLOalpha}, 
depending on what jet reconstruction algorithm is used.

\begin{figure}[bt]
\hfil
\includegraphics[height=1.6truein,keepaspectratio=true]{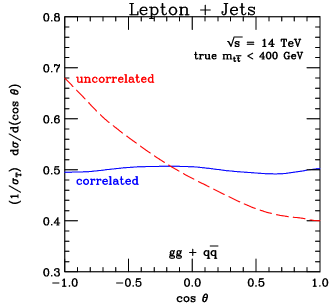}
\hfill
\caption[]{The differential distribution in $\cos\theta$,
$(1/\sigma_{{}_T}) \drm\sigma/\drm(\cos\theta)$, where $\theta$
is the ZMF angle between the charged lepton and the $d$-quark jet
(defined to be the jet which is spatially closest to the $b$-tagged
jet in the $W$ rest frame; this is also the jet with the lowest energy
in the top quark rest frame).  The solid curve is for the fully-correlated
case whereas the dashed curve assumes that the top quarks decay
spherically in their respective rest frames.  A cut restricting the
invariant mass of the $t\bar{t}$ pairs to a maximum of 400~GeV
has been applied to these distributions.
}
\label{fig:cth}
\end{figure}

Because the ZMF is well-determined in lepton+jets mode, it is possible
to examine the actual opening angle between the lepton and the $d$-jet
candidate (Fig.~\ref{fig:cth}).  
\noindent
Taking half of the area between
the correlated and uncorrelated curves to be a measure of the
potential sensitivity of this observable, we note that this half-area
is equal to 0.07 for the $\cos\theta_{\bar{e}d}$ distribution, 
as compared to a value of 0.11 for the $\Delta\phi_{\bar{e}\mu}$ 
distribution.  However, this somewhat reduced sensitivity  could
easily be offset by the higher statistics of the lepton+jets mode.
Furthermore, the systematics of this measurement are certainly very
different than those associated with the $\Delta\phi$ distribution;
thus, further investigation of this variable by the experimental 
collaborations is warranted.

\begin{figure}[hbt]
\hfil\includegraphics[height=1.6truein,keepaspectratio=true]{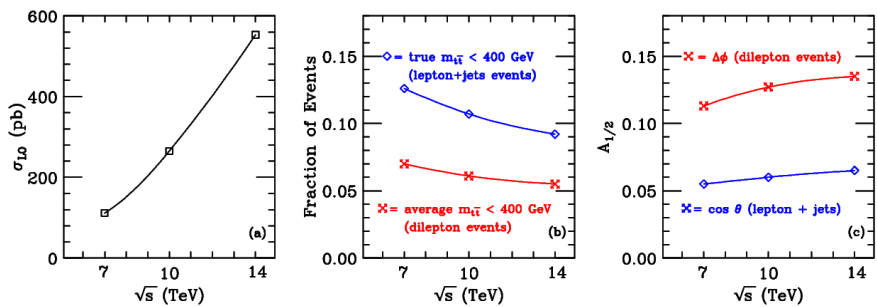}
\caption[]{Effects of varying the machine center-of-mass energy
$\sqrt{s}$. (a) Total leading order cross section for 
$pp\rightarrow{t\bar{t}}$.  These values should be multiplied by 
the branching fraction to dileptons (4.6\%) or lepton-plus-jets
(29\%), as appropriate.  We include only the $e$ and $\mu$ channels.
(b) Fraction of dilepton and lepton plus jets events with 
$m_{t\bar{t}}<400$~GeV.  For dilepton events (crosses) we employ
the unweighted average of the up to 8 solutions for the $t\bar{t}$
invariant mass.  For lepton+jets events (diamonds) the true value
of $m_{t\bar{t}}$ may be reconstructed and used in event selection.
(c) Half of the area between the appropriate unit-normalized angular
distributions for the fully correlated and spherical cases.  For
leptons+jets events (crosses), we use the distribution in $\cos\theta$,
where $\theta$ is the angle between the charged lepton and the $d$-jet
candidate in the zero momentum frame of the event.  For dilepton
events (diamonds), we use the azimuthal opening angle $\Delta\phi$
between the two charged leptons.
}
\label{fig:energydep}
\end{figure}

Before closing this section on the LHC, we present 
Fig.~\ref{fig:energydep}, which summarizes the effect of
running the LHC at reduced energy on the ability to observe
$t\bar{t}$ spin correlations.  The conclusion to be drawn from 
the three plots in Fig.~\ref{fig:energydep} is that the biggest
penalty of running at reduced energy comes from the much
smaller $t\bar{t}$ production cross section:  the size of the 
spin correlations is essentially unchanged as one reduces
$\sqrt{s}$ from 14~TeV to 7~TeV.  Thus, the prospects of observing
these correlations with the data collected during the early stages
of running are good.

\noindent

\section{The future}\label{sec:future}
Before concluding, let us take a moment or two to comment on the
sorts of precision tests of the Standard Model 
that could be done with the few million $t\bar{t}$
pairs that will ultimately be available at the LHC, beginning
with the couplings between gluons and the top quark.
All 3 of the diagrams for $gg\rightarrow t\bar{t}$ influence
the spin correlations;
these diagrams are related in a
very specific manner to satisfy SU(3) gauge-invariance.
Thus, a test of the spin correlations can be viewed as a test of
QCD gauge-invariance.

The time scale for the top quark is expected to be much shorter
than the spin decorrelation time scale.  This fact
offers us the opportunity to perform a direct test
of the predicted $V-A$ structure of the $Wtb$ vertex, the only quark
for which such a test is possible.  
By measuring the size of the correlations for different combinations
of decay products, it will be possible, with sufficient data, to 
perform measurements of the analyzing powers of all of the
top quark decay products listed in 
Table~\ref{tab:alphas}\footnote{Since the neutrino is over-constrained
in lepton+jets mode, a measurement of $\alpha_\nu$ may be possible, and
a detailed feasibility study should be performed to investigate this
possibility.}

In addition, one ought to look for
non-Standard Model couplings and/or couplings to new particles, such as
Kaluza-Klein gluons, ``extra'' Higgs bosons, or ``extra'' $Z$-bosons.
Should a new particle that couples strongly to the top 
quark be observed as
a resonance in the $t\bar{t}$ channel, then the spin correlations
of the $t\bar{t}$ pairs produced in the resonance region of phase
space would bear the imprint of the spin and parity of the
new particle and serve as a useful diagnostic in identifying
the nature of the new physics.

\section{Summary and conclusions}\label{sec:conclusions}

In summary, we have seen that the Tevatron and LHC probe two different
aspects of spin correlations in top quark pair production and decay.
At the Tevatron, the production cross section is dominated by
$q\bar{q}\rightarrow t \bar{t}$.  For this process, the off-diagonal
spin basis provides the largest signal of spin correlations.
The strategy for observing the correlations at the Tevatron involves
extracting a joint decay distribution utilizing decay angles 
in the $t$ and $\bar{t}$ rest frames.  At present, this challenging
measurement is limited by low statistics.

At the LHC, the production cross section is dominated by 
$gg\rightarrow t \bar{t}$.  
At high values of the $t\bar{t}$ invariant mass,
opposite helicity gluons to opposite spin $t\bar{t}$ pairs dominate
the cross section;  the correlations involved are the same as
for $q\bar{q}\rightarrow t\bar{t}$.  At low values of
the $t\bar{t}$ invariant mass, like helicity gluons dominate the
cross section; these gluons primarily 
produce like-helicity $t\bar{t}$ pairs.  By cutting on the value
of $m_{t\bar{t}}$, it is possible to enhance the contributions from
either like or opposite helicity gluon pairs.  In addition to a 
repeat of a Tevatron-style analysis, promising observables for
observing spin correlations include the azimuthal opening angle
between the two leptons in dilepton events and the cosine of the
angle between the lepton and $d$-jet candidate in the 
ZMF in lepton+jets mode.  With millions of $t\bar{t}$ pairs
on the horizon, precision (\%-level) measurements of the various
correlation parameters should be possible during the next decade.

\acknowledgments
The author would like to thank Stephen Parke for providing
his insight and
unique perspective on many of the topics associated with this work.
Funding to present this talk in Bruges was provided by 
the Office of the Director of Academic Affairs
at Penn State Mont Alto,
the Penn State Mont Alto Faculty Affairs Committee Professional
Development Fund,
and the
Eberly College of Science Professional Development Fund.

\end{document}